\documentclass[conference,a4paper]{IEEEtran} 
\usepackage[linesnumbered,ruled,vlined]{algorithm2e}
\usepackage{amsmath}
\usepackage{amssymb}
\usepackage{tabularx}
\usepackage{multicol}
\usepackage{wrapfig}
\usepackage{amssymb}
\usepackage{siunitx}
\usepackage{multirow}
\usepackage{stfloats}
\usepackage{xcolor}
\usepackage{graphicx}
\usepackage{algorithmic}
\usepackage{floatrow}
\usepackage{url}
\usepackage{floatrow}
\ifCLASSINFOpdf

\fi

\begin{document}

\title{Similarity Learning based Few Shot Learning for ECG Time Series Classification}

\author{\IEEEauthorblockN{Priyanka Gupta}
\IEEEauthorblockA{\textit{CSIS Department} \\
\textit{BITS-Pilani,Hyderabad Campus}\\
\textit{(CVR College of Engineering)}\\
Hyderabad, India \\
p20190501@hyderabad.bits-pilani.ac.in}

\and
\IEEEauthorblockN{Sathvik Bhaskarpandit}
\IEEEauthorblockA{\textit{CSIS Department} \\
\textit{BITS Pilani Hyderabad}\\
Hyderabad, India \\
f20191200@hyderabad.bits-pilani.ac.in}
\and
\IEEEauthorblockN{Manik Gupta}
\IEEEauthorblockA{\textit{CSIS Department} \\
\textit{BITS Pilani Hyderabad}\\
Hyderabad, India \\
manik@hyderabad.bits-pilani.ac.in}
}


\maketitle

\begin{abstract}

  Using deep learning models to classify time series data generated from the Internet of Things (IoT) devices requires a large amount of labeled data. However, due to constrained resources available in IoT devices, it is often difficult to accommodate training using large data sets. This paper proposes and demonstrates a Similarity Learning-based Few Shot Learning for ECG arrhythmia classification using Siamese Convolutional Neural Networks. Few shot learning resolves the data scarcity issue by identifying novel classes from very few labeled examples. Few Shot Learning relies first on pretraining the model on a related relatively large database, and then the learning is used for further adaptation towards few examples available per class.Our experiments evaluate the performance accuracy with respect to K (number of instances per class) for ECG time series data classification. The accuracy with 5- shot learning is 92.25\% which marginally improves with further increase in K. We also compare the performance of our method against other well-established similarity learning techniques such as Dynamic Time Warping (DTW), Euclidean Distance (ED), and a deep learning model - Long Short Term Memory Fully Convolutional Network (LSTM-FCN) with the same amount of data and conclude that our method outperforms them for a limited dataset size. For K=5, the accuracies obtained are 57\%, 54\%, 33\%, and 92\% approximately for ED, DTW, LSTM-FCN, and SCNN, respectively.

\end{abstract}

\IEEEpeerreviewmaketitle

\section{Introduction}

Timeseries are a significant class of temporal data objects. It is generated effortlessly by recording a  sequence of observations chronologically. Examples of time series are the price of stocks,  electrocardiogram  (ECG), target star's brightness,  daily temperature,  weekly sales totals, etc \cite{fu2011review}. Time series classification  (TSC)  task is training a  classifier on a  dataset $D =  \{(X_{1}, Y_{1}),(X_{2}, Y_{2}), . . . ,(X_{N} , Y_{N} )\}$ to map inputs $X_{i}$ into one of the existing classes $Y_{i}$.  Here, $X_{i}$ is a  univariate or multivariate time series, and $Y_{i}$ is the class label. 

ECG classification comes under Time series classification (TSC) problem. Studies show that the number of mortalities will increase from 17.3 million (2016) to 23.6 million (2030) \cite{american2016heart} due to prevalence of cardiovascular diseases. ECG being noninvasive and painless is often used as a preliminary step towards treating heart diseases. ECG classification is important as it provides valuable information about the functioning of the circulatory system. ECG is a univariate time series that has been used as the input for our work and consists of five different classes. 

 TSC tasks are different from traditional classification problems as the attributes have a temporal ordering; however, various machine learning and deep learning-based techniques have been applied in the literature to address this problem. In \cite{bagnall1602great} authors have evaluated 18 machine learning-based time series algorithms and concluded Dynamic Time Warping (DTW) performs better after Collective of Transformation Ensembles (COTE) ensembles. Similarly, \cite{fawaz2019deep} presents a study of 9 different deep learning techniques and concludes Residual Network (ResNet) followed by Fully Convolutional Network (FCN) performs better among them. However, the performance for each of these techniques is constrained by the amount of labeled data.

With the current methodology, we need a subject matter expert (SME) to label data, and then the supervised learning techniques are applied to it. Manual labelling may result in inexact and noisy labels. Also, the cost of labelling the data and the time required for big data labelling is very high and not scalable. Therefore, there is a need for learning techniques that can do cost-effective data classification with few labelled data and have accuracy at par with the supervised learning techniques that use large labelled data sets. From our study, we infer that there are research gaps in the design and development of time series classification techniques with few labelled data.

The explosive growth of the Internet of Things (IoT) devices emphasizes the need to develop classification methodologies that can generalize well from few labelled data. IoT devices are generally resource-constrained and cannot store large data sets for training purposes. There are high costs associated with obtaining labelled data as well. Further, the properties of ECG data collected by IoT devices also present specific challenges like (1) Class imbalance problem due to rare occurrence of some ECG events (2) Low sampling frequency, single ECG lead and unavoidable noisiness may cause low signal quality and (3) Small labelled data set due to high cost of labelling, and non-scalability \cite{weimann2021transfer}. In the current work, we try to address the first and third challenges by applying Siamese Convolutional Neural Network (SCNN) for class imbalance problems and few-shot learning (FSL) for unavailability of extensive labelled data.

In this work, we specifically try to address the problem of ECG classification using few labelled examples and this approach is termed Few Shot Learning. FSL is a Weak Supervised Learning (WSL) technique with incomplete supervision where only a small amount of samples have supervised information \cite{wang2020generalizing}. FSL uses a transfer learning approach \cite{sun2019meta} that tries to solve the task of learning from a few given examples, called the support set. Transfer learning is a technique wherein knowledge gained in learning for one task is utilized for solving another task \cite{olivas2009handbook}. In FSL, the model is pre-trained on a different related large dataset and then uses the learning for a new task with fewer labelled data samples per class. FSL techniques are efficient in bridging the gap between model generalization performance and data set size \cite{triantafillou2017few}.

Previous works have shown that transfer learning shows promising results for solving the TSC \cite{kashiparekh2019convtimenet, fawaz2018transfer, kimura2020convolutional}. It involves the idea of pre-training a model on large data to prevent overfitting, then tuning the pre-trained model to classify new data. In \cite{kashiparekh2019convtimenet} authors selected 24 UCR datasets to perform pre-training and tested on another 41 data sets from UCR for generalizability and transferability. Their experiment showed a significant gain in classification accuracy and computational efficiency. In \cite{fawaz2018transfer} authors have carried out transfer learning for each pair of UCR time-series datasets and shown the dependence of the performance of a model on a pre-training dataset. Further they have given a DTW based solution for selecting the pre-training dataset. In \cite{kimura2020convolutional} authors have used a CNN based transfer learning for flood prediction. They have shown CNN with transfer learning can effectively shorten the training time by $1/5^{th}$ and a mean error difference by 15\% as compared to CNN without transfer learning. However, few shot learning for ECG data classification using SCNN is not yet explored.. 

In this work, we demonstrate the application of Siamese networks for few shot classifications of ECG. The SCNN is used to make predictions on the query set using the support set, which consists of $K$ examples per class. 
SCNN \cite{koch2015siamese} learns a similarity score between pairs of time series. Since SCNN makes use of pairs of time series to learn similarity score, it is not affected by class imbalance \cite{malialisa2020data}, and this is one of the reasons for using SCNN in our work. In particular, we use a CNN with various filters and concatenate them to learn a feature embedding of each time series.

The key contributions of this work are summarised as follows:
\begin{enumerate}
  \item We demonstrate Siamese Convolutional Neural Network (SCNN) for few shot time series classification.
  \item We demonstrate that the SCNN can learn effective feature embeddings by using one-dimensional filters of various lengths.
  \item We demonstrate that the SCNN can adapt to any $N$ way $K$ shot learning task using a smaller number of training samples from the target task. We observe that few shot classifications using SCNN varies marginally after $K$=5 for ECG classification. 
\end{enumerate}

The rest of the paper is structured as follows: Section \ref{Related_work} presents the related work done in the area of ECG classification and few-shot learning techniques for time series data. Section \ref{Methodology} gives the details of the methodology used, Section \ref{Dataset} provides datasets, and experimental details, Section \ref{Results} provide details on the results obtained and comparison with baseline methods. Finally, the conclusions and future work are in Section \ref{Conclusion}.

\section{Related work}
\label{Related_work}
This section presents a literature review of the current state of the art for ECG classification techniques and few shot learning for time series data. First subsection \ref{ECG_cf_tech} presents related work for ECG classification techniques. Following subsection \ref{FSL_tech_ts} describes few-shot learning for time series data.

\subsection{ECG classification techniques}
\label{ECG_cf_tech}

Electrocardiogram (ECG) signals have been used significantly for the diagnosis of heart diseases.  Traditional methods for ECG classification used by clinicians are usually strenuous, time-consuming, and susceptible to clinical experience variability. Current research trends on deep learning show that models trained on a huge amount of data can carry out feature extraction directly from the labelled data and diagnose cardiac arrhythmia better than professional cardiologists. \cite{8759878}.
 
 ECG classification studies started with the application of Genetic Algorithms, and Machine Learning algorithms \cite{plawiak2018novel}, \cite{pplawiak2018novel} that focused on designing methods for ECG pre-processing, feature extraction, feature selection, training and testing. With the availability of high computational hardware, deep learning models gained a research focus. Deep learning models provide an advantage of automated feature extraction. Several DL models such as two-dimensional CNNs \cite{weimann2021transfer}, Long Short Term Memory (LSTMs) \cite{yildirim2018novel}, Deep Belief Networks (DBN) \cite{song2019automatic}, one-dimensional Convolutional Neural Networks (CNNs) \cite{acharya2017automated},  have been proposed that achieve state of the art results on TSC datasets as well as ECG classification.  In \cite{weimann2021transfer} authors have used two dimensional CNN and showed that using a transfer learning approach that involves pre-training on a large database. Then fine-tuning on a relatively small database improves the performance of the target task. Authors have used  Icentia11K (largest ECG data set) for pre-training and PhysioNet/CinC Challenge 2017 data set for fine-tuning. In \cite{yildirim2018novel} authors have used Deep Unidirectional LSTM wavelet sequence (DULSTM-WS) and Deep Bidirectional LSTM wavelet sequence (DBLSTM-WS) to get an accuracy of 99.25\% and 99.39\%. In \cite{song2019automatic} authors have used discriminative DBNs and achieved an accuracy of 98.49\%. Higher accuracy is seen in \cite{yildirim2018novel,song2019automatic} because the training set and test set have been taken from the same database. In \cite{acharya2017automated} authors have taken two seconds and five-second segments of ECG and applied one dimensional CNN that results in the accuracy of 92.50\% and 94.9\%. They have used multiple datasets (MIT-BIH Atrial Fibrillation (afdb) \cite{goldberger2000physiobank}, MIT-BIH Arrhythmia (mitdb) \cite{moody1983new}, Creighton University Ventricular Tachyarrhythmia (cudb) \cite{nolle1986crei}) for better generalization. However, in general, most of the techniques available in the literature are constrained by the availability of large labelled data. Therefore, techniques that require fewer labelled samples for ECG classification need to be explored, and towards that direction, we endeavour to explore FSL techniques in the next subsection.

\subsection{FSL techniques for time series data}
\label{FSL_tech_ts}

Few shot learning is used to learn from scarce data by identifying novel classes from very few labelled examples \cite{geng2019induction}.  The seminal research work on few-shot learning started in the early 2000s \cite{fe2003bayesian, fei2006one}. The authors used generative models together with complex iterative inference strategies. Few-shot classification has been widely applied thereafter in various application as object recognition\cite{fei2006one}, image classification \cite{vinyals2016matching}, and text sentiment classification \cite{yu2018diverse}. In a recent survey paper on FSL, \cite{wang2020generalizing} authors have analyzed a few shot learning from three perspectives, i.e. data, model and algorithm.

In \cite{narwariya2020meta} authors have used few shot time series classification as a meta-learning approach for UCR data sets belonging to various domains. The experiments have shown that few shot time series classifications can classify a target domain using only a smaller number of training samples from the target task. The applicability of meta-learning approaches is limited as they are biased towards the fixed number of target classes across tasks. Using few-shot learning, they overcome this limitation to train a common agent across domains, with each domain having a different number of target classes. This approach of classification outperforms Euclidean distance (ED), Dynamic time wrapping (DTW), Bag-of-SFA-Symbols (BoSS), Residual network (ResNet). However, our work differs from them as they have done pre-training on various domains (e.g. healthcare, activity recognition, etc.), which gives lesser classification accuracy. We are using different ECG datasets for pre-training and fine-tuning that results in better classification accuracy.

In \cite{guptatime} they have used Reptile with FCN for time series classification with meta-learning. The experiment on the univariate UCR dataset shows that few-shot learning leads to faster convergence with fewer iterations over the non-meta-learning equivalent.

In our work, we have demonstrated the application of a few-shot learning method for ECG classification that can work with limited resources in memory, power, and time. We need to store a few labelled data instead of a large dataset for training, so FSL uses less memory than traditional supervised learning. Since less data means less time and power requirement, so can be used in wearable devices.

\section{Methodology}\label{Methodology}
\begin{figure*}[htp]
    \centering
    \includegraphics[width= 0.8\linewidth]{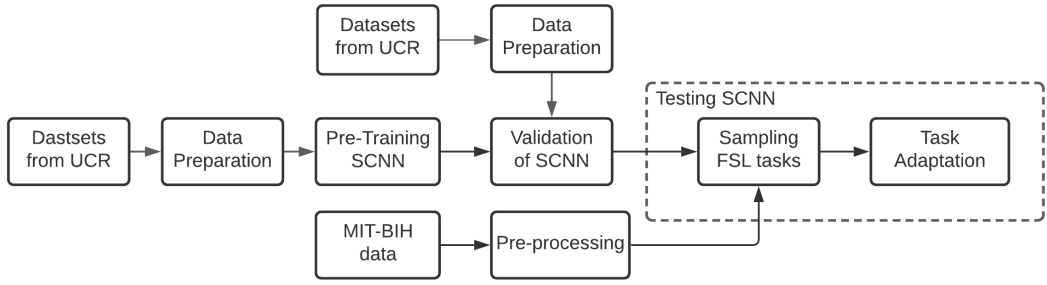}
    \caption{Overview of FSL using SCNN}
    \label{fig:flowchart}
\end{figure*}

This section presents our method of using similarity learning to classify ECG data under the FSL scenario accurately. Figure \ref{fig:flowchart} shows a flowchart of the various steps involved. We summarize the major steps taken as follows:

\begin{enumerate}

    \item Data Preparation: We make pairs from the training and validation sets and assign their corresponding labels.
    \item Pre Training: We utilize the training pairs and their labels that are prepared from training datasets selected from UCR Archive \cite{UCRArchive} to train the Siamese Network.
    \item Validation: We utilize the validation pairs and their labels prepared from validation datasets selected from UCR Archive, not used in training, for validation and tuning of hyperparameters of the Siamese Network.
    \item Task Sampling: We preprocess data from the MIT-BIH database and sample $N$ way $K$ shot learning tasks from the processed data.
    \item Task Adaptation: We utilize our pre-trained Siamese Network to obtain predictions for each task and record the accuracy using the predicted labels and the actual class labels.
    
\end{enumerate}

\subsection{Data Preparation}
Once the datasets have been split into training, validation, and test sets, denoted by $D_{tr}, D_{val}$ and $D_{test}$ respectively, they have to be preprocessed to learn similarity efficiently. 
Each set contains number of time series and their labels which can be represented as a set $D = \{ (X_{i}, Y_{i}) | i = 1, 2, 3, ..., M \}$ where M is the number of time series in the set, $X_{i}$ is a time series or a set of samples at different time stamps of the form $\{ x_{1}, x_{2}, ... x_{T} \}$, T being the number of timestamps or time series length of a particular time series, and $Y_{i}$ is the class label corresponding to the time series $X_{i}$.

To make the data in the training and validation sets ready to be fed into the Siamese Network, we perform the following steps. Each time series from $D_{tr}$ and $D_{val}$ is scaled to the same range and post-padded with zeros \cite{dwarampudi2019effects} to match to the same length of $L_{max}$, which is the length of the longest time series among all the three datasets. From $D_{tr}$ we create training pairs $P_{tr}$ and from $D_{val}$ we create validation pairs $P_{val}$ using Algorithm \ref{Algorithm 1}. Each pair from $P_{tr}$ and $P_{val}$ is of the form $( (X_{1}, X_{2}) , Y')$ where $X_{1}, X_{2} \in D_{tr}$ or $X_{1}, X_{2} \in D_{val}$, and 
\begin{math}
Y' = 
\begin{cases}
    1 & \mbox{if }  y(X_{1}) = y(X_{2})  \\
    0 & \mbox{if }  y(X_{1}) \neq y(X_{2})  \\
\end{cases}
\end{math}
where $y(X_{i})$ is the class label $Y_{i}$ corresponding to a time series $X_{i}$ from $D_{tr}$ or $D_{val}$.

\begin{algorithm}
\SetAlgoLined\SetArgSty{}
\DontPrintSemicolon
    \For {dataset in set $D_{tr}$}    
    { 
        \For{class label Y in the list of dataset's unique class labels}
        {
            $n$ = number of time series with label Y\;
            $p = \binom{n}{2} = \frac{n \times (n-1)}{2}$\; \tcp{Number of pairs p given n}
            Create $p$ 'same label' pairs of the form $( (X_{1}, X_{2}) , Y')$ from the n time series\;
            Set $Y'=1$ for the above p pairs \; 
            Append the above pairs to $P_{tr}$\;
            Randomly sample $p$ time series having label Y\;
            Randomly sample $p$ time series not having label Y\; 
            Create $P$ 'different label' pairs of the form $( (X_{1}, X_{2}) , Y')$\;
            Set $Y'=0$ for the above p pairs \;
            Append the above pairs to $P_{tr}$\;
        }
    }
    Repeat steps 1 to 14 for $D_{val}$, appending pairs to $P_{val}$\;
    
\caption{Generating Pairs for Siamese Network}
\label{Algorithm 1}
\end{algorithm}

It is to be noted that the pairs for training, validation, and final tasks for testing are sampled from disjoint sets of classes. This ensures the generalizability of our model, i.e., the ability to classify new unseen examples on previously unseen datasets. 
In practice we guarantee this by using different datasets for each of the three sets $D_{tr}$, $D_{val}$ and $D_{test}$.

\subsection{Siamese Network Pre Training and Validation}

\begin{figure*}[!hbt]
  \centering
  \label{}
  \includegraphics[width=0.7\linewidth]{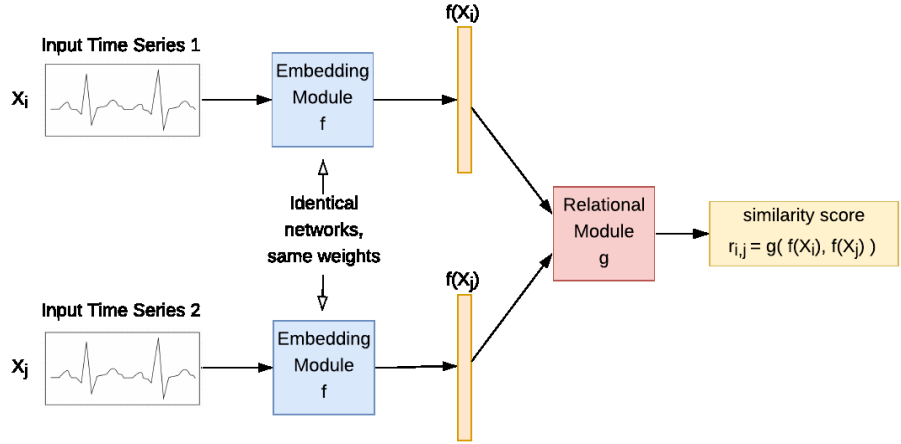}
  \caption{Structure of the Siamese Network wherein inputs are given in form of two time series and a similarity score between 0 and 1 is outputted}
  \label{fig:Siamese Network Architecture}
\end{figure*}

Once pairs for training and validation are made, the Siamese Network is ready to be trained. $P_{tr}$ is used to train the network and arrive at the optimum set of weights, while $P_{val}$ is used to tune hyperparameters and test the model's generalizability. The network architecture and working can be broadly described as follows. It inputs a pair of time series and outputs their similarity score. It consists of two modules: an embedding module followed by a relational module as shown in Figure \ref{fig:Siamese Network Architecture}.

\subsubsection{Embedding Module}

\begin{figure*}[htp]
  \centering
  \includegraphics[width= \linewidth]{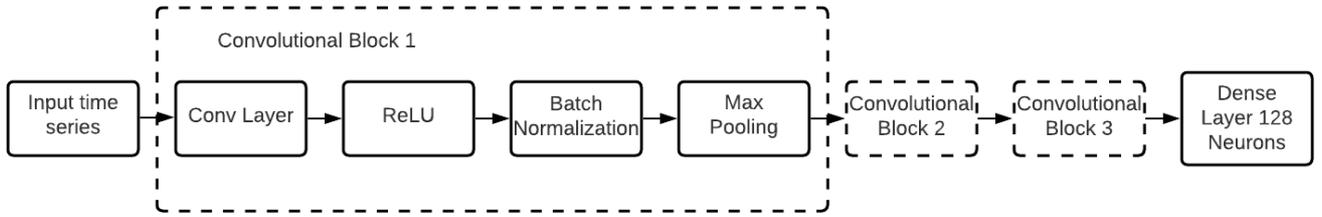}
  \caption{Embedding Module Architecture}
  \label{fig:Embedding Module}
\end{figure*}

The embedding module computes an embedding or representation $z = f(X)$ of each time series X. It consists of a 1-D convolutional neural network (CNN). The input is a padded time series $X$. There are three main convolutional blocks. Each convolutional block consists of a filter and Rectified Linear Unit (ReLU) as its activation function. It is followed by a batch normalization operation and a dropout layer that act as regularizers, and finally, a pooling layer. The first convolutional block consists of 128 convolutional filters of length 7 and a pooling layer of size 3. The second convolutional block consists of 64 filters of length 5 and a pooling layer of size 3. The third and last convolutional block consists of 64 filters of length 5 and a pooling layer of size 2. Various filters of multiple lengths are used in each convolutional layer to extract useful temporal information from time series of varying lengths at different time scales. The pooling layers help in reducing the dimensions of the feature maps produced by the filters and summarize the features present in a region of the feature maps.  Figure \ref{fig:Embedding Module} shows the overall architecture of the embedding module.

\subsubsection{Relational Module}
The relational module calculates a similarity score between two time series $X_{i}$ and $X_{j}$ given their embeddings $f(X_{i})$ and $f(X_{j})$, i.e. , it computes  
\begin{equation}
    r_{i,j} = g(f(X_{i}), f(X_{j})) \\
\end{equation}
The function $g$ is taken to be a symmetric function so that the similarity score is independent of the order of the inputted time series. The exact equation for the similarity score is as follows:
\begin{equation}
    r_{i,j} = \sigma(| f(X_{i}) - f(X_{j}) |)
\end{equation}
 where $\sigma$ represents the sigmoid function. Thus, the similarity score is a value between 0 and 1, with a higher value representing a higher degree of similarity between the two time series.

\subsection{Testing the Siamese Network}

Given the preprocessed test set, we test our Siamese Network by first sampling FSL tasks and then making the network adapt to each task to give final predictions. Each task consists of randomly sampling $K$ time series from each of the $N$ classes, that act as the support set, then randomly sampling $Q$ time series from each of the $N$ classes, that act as the query set. Our objective is to accurately classify the $NQ$ queries given $NK$ samples. To make our Siamese Network adapt to any such $N$ way $K$ shot task, we do the following step. From the embedding module, the embeddings of the support set $f(X_{i})$, i $\in {1, 2 , ..., NK}$ are obtained. These embeddings are then averaged class-wise to obtain $N$ feature vectors $\mu_{1}, \mu_{2}, ... , \mu_{N}$, each representing one of the N classes. The queries' embeddings $Q_{j} = f(X_{j})$, j $\in {1, 2 , ..., NQ}$ are also obtained. 
The final class label for each query is given as:
\begin{equation}
  \hat{y_{j}} = \operatorname*{argmax}_i \{g(\mu_{i}, Q_{j}) | i = 1...N\}  
\end{equation}

\section{Datasets and Experimental Details}
\label{Dataset}

\subsection{Dataset Description}
In this paper, we have used the ECG200, and ECG5000 datasets from the UCR Archive \cite{UCRArchive} for training purposes, TwoLeadECG and ECGFiveDays datasets from the UCR Archive \cite{UCRArchive} for validation. We have selected different data sets for training and validation to decrease overfitting and generalization errors. 

\begin{table}[hbt!]
\caption{UCR Archive Datasets' Characteristics}
\label{ds_desc}
    \begin{tabular}{|c|c|c|c|}
 
        \hline
            Dataset Name & Size & Length & No. of Classes\\[0.5ex]
        \hline
            ECG200 & 200 & 96 & 2 \\
            ECG5000 & 5000 & 140 & 5 \\
            ECGFiveDays & 884 & 136 & 2 \\
            TwoLeadECG & 1162 & 82 & 2 \\
            MIT-BIH & 21892 & 187 & 5\\
        \hline
    \end{tabular}
\end{table}

The testing for few shot classification has been carried out on MIT-BIH Arrhythmia Database. \cite{https://doi.org/10.13026/c2f305}. The MIT-BIH dataset has ECG recordings from 47 different subjects recorded at the sampling rate of 360Hz. Two cardiologists have labeled each beat independently. Further, these annotations are used to create five different beat categories denoted as N, S, V, F, and Q. The properties of the different ECG datasets used are shown in Table \ref{ds_desc}. The maximum length of time series in MIT-BIH dataset is 187.





 


\subsection{Experimental Setup}

\subsubsection{Pretraining and Validation} 
Time series from various datasets of the UCR Archive are first scaled to values between 0 and 1 and then padded with zeros to a length $L_{max}$ of 187. Pairs are made using the Algorithm \ref{Algorithm 1}. The pretraining then becomes a binary classification problem with the objective of classifying a pair of time series correctly. Weights are taken so as to minimize the binary cross-entropy of the validation set between the binary class label and the predicted similarity score. The Adam optimizer \cite{kingma2017adam} is used with an initial learning rate of $10^{-3}$ and early stopping with the patience of 20 epochs. The optimum number of filters, filter sizes, and pooling sizes is found out using the Hyperband algorithm \cite{li2017hyperband} to minimize the validation loss.

\subsubsection{FSL Task Sampling and Adaptation}
\label{samp_FST}
Since there are 5 classes in the test set (MIT-BIH dataset), each task sampled forms a 5-way $K$-shot learning problem whose accuracy needs to be determined. For each FSL task, we randomly sample K time series from each of the 5 classes, which forms the support set. We also randomly sample 20 queries per class, giving a total of 100 queries per task. From the embedding module, we get $5K$ embeddings for the support set and $5Q$ embeddings for the query set. From the embeddings of the support set, we obtain 5 feature vectors via class-wise averaging of $K$ embeddings. For each query, we feed its embedding along with one of the feature vectors into the relational modules to obtain a similarity score and repeat 5 times with each of the feature vectors. The final predicted label for the query is taken as the maximum of the 5 similarity scores. The above process is repeated for each of the 100 queries, and the accuracy of the task is measured. The final metrics, such as accuracy, macro precision, macro recall, and macro F1 score, are reported as the mean of the corresponding metrics of 20 such tasks.

\subsection{Baselines Considered}

We consider the following baselines and compare their results with our model.

\subsubsection{Dynamic Time Warping (DTW)} 
Nearest Neighbour based on Dynamic Time Warping (1NN-DTW) has been shown to work well for univariate time series classification (UTSC) \cite{bagnall1602great}. An advantage of DTW is that no preprocessing is required, even for time series of different lengths. The final accuracy is reported as the average of 20 tasks. Tasks are sampled in the same way as for our Siamese Network, as mentioned in Section \ref{samp_FST}. For every query, DTW distances are computed between the query and every element in the support set, and the label of the nearest Neighbour is taken as the label of the query.

\subsubsection{Euclidean Distance (ED)}
1-NN based on Euclidean Distance is one of the oldest, simplest, and computationally efficient methods for UTSC. However, it has been shown to perform inferior to DTW \cite{ratanamahatana2004making} and has the additional condition that the time series have to be of the same length. Here we report the final accuracy as average accuracy of 20 tasks of 100 queries each.

\subsubsection{LSTM-Fully Convolutional Network (LSTM-FCN)}

The deep learning baseline used is the LSTM-FCN as suggested in \cite{karim2017lstm}, which has been shown to outperform recurrent or convolutional models. It entails the augmentation of fully convolutional networks with LSTM submodules for TSC. We train the model for 200 epochs and use an Adam optimizer with an initial learning rate of $10^{-4}$. The experiment is repeated twenty times, and the average accuracy is reported.

\section{Results and Discussion}

\begin{table}[t]

\caption{Accuracy of various models with varying $K$}
\label{Accuracy Table}
\begin{tabular}{|c|c|c|c|c|}
\hline
\multirow{2}{*}{$K$} & \multicolumn{4}{c|}{Accuracy}\\ \cline{2-5} 
                   & ED     & DTW    & LSTM-FCN & SCNN\\ \hline
1                  & 0.4280 & 0.3505 & 0.3475   & 0.8475 \\
2                  & 0.4835 & 0.4010 & 0.3375   & 0.9050 \\
3                  & 0.5245 & 0.4600 & 0.3215   & 0.9195 \\
4                  & 0.5535 & 0.5215 & 0.3220   & 0.9195 \\
5                  & 0.5715 & 0.5495 & 0.3350   & 0.9225 \\
10                 & 0.6340 & 0.6065 & 0.3355   & 0.9190 \\
20                 & 0.7020 & 0.6255 & 0.6385   & 0.9205 \\
30                 & 0.7190 & 0.7145 & 0.7765   & 0.9210 \\
40                 & 0.7450 & 0.7390 & 0.7780   & 0.9210 \\
50                 & 0.7645 & 0.7705 & 0.8215   & 0.9225 \\ \hline
\end{tabular}
\end{table}

\begin{figure}[t]
  \centering
  \includegraphics[width=\linewidth]{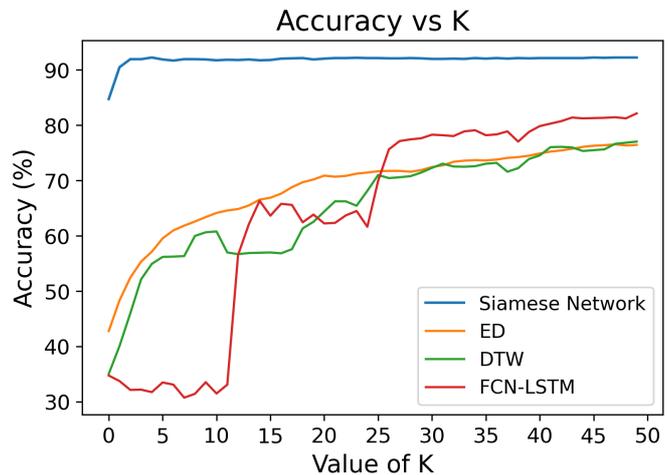}
  \caption{Comparison of various models with their average accuracies plotted versus $K$}
  \label{mult:accuracy}
\end{figure}

\begin{table*}[t]
\centering
\caption{Different metrics of various models with varying $K$}
\label{Metrics Table}
\begin{tabular}{|c|c|c|c|c|c|c|c|c|c|c|c|c|} 
\hline
\multirow{2}{*}{$K$} & \multicolumn{4}{c|}{Precision}                                            & \multicolumn{4}{c|}{Recall}                                               & \multicolumn{4}{c|}{F1 Score}                                              \\ 
\cline{2-13}
                   & ED    & DTW   & \begin{tabular}[c]{@{}c@{}}LSTM-\\FCN\end{tabular} & SCNN  & ED    & DTW   & \begin{tabular}[c]{@{}c@{}}LSTM-\\FCN\end{tabular} & SCNN  & ED    & DTW   & \begin{tabular}[c]{@{}c@{}}LSTM-\\FCN\end{tabular} & SCNN   \\ 
\hline
1                  & 0.446 & 0.414 & 0.569                                             & 0.857 & 0.428 & 0.351 & 0.348                                             & 0.848 & 0.437 & 0.379 & 0.432                                             & 0.852  \\
2                  & 0.499 & 0.445 & 0.616                                             & 0.909 & 0.484 & 0.401 & 0.338                                             & 0.905 & 0.492 & 0.442 & 0.436                                             & 0.907  \\
3                  & 0.536 & 0.515 & 0.629                                             & 0.923 & 0.525 & 0.460 & 0.322                                             & 0.919 & 0.530 & 0.486 & 0.425                                             & 0.921  \\
4                  & 0.557 & 0.537 & 0.632                                             & 0.924 & 0.554 & 0.522 & 0.322                                             & 0.919 & 0.555 & 0.529 & 0.427                                             & 0.922  \\
5                  & 0.569 & 0.554 & 0.643                                             & 0.926 & 0.572 & 0.549 & 0.335                                             & 0.923 & 0.570 & 0.552 & 0.441                                             & 0.924  \\
10                 & 0.632 & 0.616 & 0.658                                             & 0.923 & 0.634 & 0.607 & 0.336                                             & 0.919 & 0.633 & 0.611 & 0.445                                             & 0.921  \\
20                 & 0.704 & 0.628 & 0.736                                             & 0.924 & 0.702 & 0.626 & 0.639                                             & 0.921 & 0.703 & 0.627 & 0.684                                             & 0.923  \\
30                 & 0.720 & 0.721 & 0.794                                             & 0.925 & 0.719 & 0.715 & 0.777                                             & 0.921 & 0.720 & 0.718 & 0.785                                             & 0.923  \\
40                 & 0.746 & 0.739 & 0.803                                             & 0.925 & 0.745 & 0.739 & 0.778                                             & 0.922 & 0.746 & 0.739 & 0.796                                             & 0.923  \\
50                 & 0.768 & 0.775 & 0.832                                             & 0.926 & 0.765 & 0.771 & 0.822                                             & 0.923 & 0.766 & 0.773 & 0.827                                             & 0.924  \\
\hline
\end{tabular}
\end{table*}

\label{Results}

We evaluate results based on the average accuracies of the tasks sampled as mentioned in section \ref{samp_FST}. We vary the value of $K$ and observe its effect on the final accuracy. Table \ref{Accuracy Table} shows the accuracy obtained using different models with selected values of $K$, while Figure \ref{mult:accuracy} shows the plot of accuracy for the various models as $K$ is varied from 1 to 50.  



The plots and tables show some interesting results. ED follows a nearly monotonically increasing behavior of precision, recall, and F1 score with an increase in $K$. DTW does not follow such a smooth behavior and underperforms in comparison to ED for smaller values of $K$. However, we see DTW outperforms ED for values closer to 50 and could possibly do even better for values larger than 50. Unlike ED and DTW, the FCN-LSTM shows extremely irregular behavior as we keep adding samples to train, with large spikes in accuracy at certain places. This could be attributed due to the stochastic nature of the optimization of the neural networks while training, as well as the lack of labeled data, to train on.

Most interestingly, in our model, we see a huge jump in accuracy from $K$=1 to $K$=2, by nearly 0.05. From $K$=2 to $K$=3 there is a jump by around 0.015. After $K$=3, there seems to be very less change in accuracy, less than 0.0001. From $K$=5, till $K$=50, the accuracy does not increase monotonically but plateaus at around 0.92. The recall, precision, and F1 score also plateau at around 0.92, as seen in Table \ref{Metrics Table}.

A plausible hypothesis for this curve given by our model is as follows. A single sample of a time series, i.e., $K$=1, is not good enough to represent the class it belongs to. As we take the mean of the embedding of more than one time series, the representation of a class label becomes more accurate, and the model gives a 'better' similarity score. However, after a certain value of $K$, the mean of $K$ such time series has already yielded a good representation of the class, and further inclusion of time series hardly makes a difference to accuracy.

For all values of $K$, our model outperforms other models in accuracy. We do notice that the LSTM-FCN shows promising performance and could potentially outperform the Siamese Network in the presence of more data. Yet, under the few-shot learning scenario, it shows heavy overfitting and cannot adapt quickly to new, unseen examples.

\section{Conclusions and Future Work}
\label{Conclusion}
This work demonstrates transfer learning based on a few shot classifications on ECG data using Siamese convolutional neural networks that can work satisfactorily without a large amount of labelled data. Siamese convolutional neural networks make use of similarity learning that helps to overcome the problem of class imbalance as ECG data are mostly imbalanced with the majority of the beats are normal, and only a few beats are abnormal \cite{malialisa2020data}. We have shown that SCNNs are capable of learning feature embeddings using one-dimensional filters of different lengths. Our experiments show that FSL using SCNN outperforms ED, DTW, LSTM based methods. It has also been shown that after a specific limit (in our case $K$=5), increasing the value of $K$ marginally improves the performance in FSL using SCNN for ECG time-series datasets. 

In the future, we will compare the performance and overheads with other few-shot learning architectures that have shown promising results in computer vision tasks. This work can be extended by converting the original one-
dimensional ECG data into the time-frequency spectrograms as the input of 2D-CNN models. By converting to time-frequency spectrograms \cite{huang2019ecg}, one can take advantage of temporal information embedded in ECG.

\bibliographystyle{ieeetr}

\bibliography{ref}

\end{document}